# On the Nature of Nonthermal Broadening of Spectral Lines Observed by IRIS

Kyuhyoun Cho[1,2], Bart De Pontieu[2,3,4], and Paola Testa[5]
[1] Bay Area Environmental Research Institute, NASA Research Park, Moffett Field, CA 94035, USA
[2] Lockheed Martin Solar & Astrophysics Laboratory, 3251 Hanover Street, Palo Alto, CA 94304, USA
[3] Rosseland Centre for Solar Physics, University of Oslo, P.O. Box 1029 Blindern, NO-0315 Oslo, Norway
[4] Institute of Theoretical Astrophysics, University of Oslo, P.O. Box 1029 Blindern, NO-0315 Oslo, Norway
[5] Harvard-Smithsonian Center for Astrophysics, 60 Garden Street, Cambridge, MA 02193, USA

## Abstract

The origin of nonthermal broadening in solar spectra is one of the long-standing questions in solar physics. Various processes have been invoked—including unresolved flows, waves, and turbulent processes—but definitive answers are lacking. To investigate the physical processes responsible for nonthermal broadening, we examine its relation with the angle between the magnetic field and the line of sight in three different closed-field regions above plage regions at different locations on the solar disk. We obtained the nonthermal width of transition-region Si IV 1403 Å spectra observed in active regions by the Interface Region Imaging Spectrograph, after subtraction of the thermal and instrumental line broadening. To investigate the dependence of the measured broadening on the viewing angle between the line of sight and magnetic field direction, we determined the magnetic field direction at transition-region heights using nonlinear force-free extrapolations based on the observed photospheric vector magnetic field taken by the Helioseismic and Magnetic Imager on board the Solar Dynamics Observatory. We found that the nonthermal broadening shows a correlation with downward motion (redshifts) and alignment between the magnetic field and the observer's line-of-sight direction. Based on the observed correlations, we suggest that velocity gradients within plasma flowing down along the magnetic field may lead to a significant portion of the observed nonthermal broadening of transition-region spectral lines in closed fields above plage regions.

*Unified Astronomy Thesaurus concepts:* Solar transition region (1532); Solar physics (1476); Solar atmosphere (1477); Solar ultraviolet emission (1533)

## 1. Introduction

The nonthermal broadening of spectral lines is the amount of excess broadening in the observed solar spectra compared to the value expected based on the thermal properties of the emitting ion (thermal broadening) and the instrumental properties (instrumental broadening). The nonthermal broadening in transition-region lines is of high interest, because it may hold important clues to the heating mechanism that is active in the solar atmosphere, as it may provide an observational signature of energy transfer from the lower solar atmosphere. After the early report by B. C. Boland et al. (1973), many previous studies have commonly found excess broadening of about a few tens of kilometers per second for transition-region lines. Several previous analyses of nonthermal broadening are reviewed in J. T. Mariska (1992) and G. Del Zanna & H. E. Mason (2018).

Nonthermal broadening can be interpreted as a cumulative effect of multiple velocity components along the line-of-sight (LOS) direction within a single detector pixel. There are several candidates for explaining what kinds of physical processes cause the multiple velocity components, e.g., acoustic or magnetohydrodynamic (MHD) waves (J. T. Mariska et al. 1978; K. G. McClements et al. 1991; K. P. Dere & H. E. Mason 1993; A. A. van Ballegooijen et al. 2011; B. De Pontieu et al. 2015), unresolved turbulent motions (G. A. Doschek & U. Feldman 1977; D. I. Pontin et al. 2020), and magnetic reconnection events (I. M. Sarro et al. 1997;

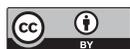 Original content from this work may be used under the terms of the Creative Commons Attribution 4.0 licence. Any further distribution of this work must maintain attribution to the author(s) and the title of the work, journal citation and DOI.

P. Testa et al. 2014, 2016, 2020; Juraj Lorincik et al. 2022; K. Cho et al. 2023). The relation between the nonthermal velocity and magnetic field direction can reveal whether the nonthermal velocity is mainly associated with plasma motions along the magnetic field or perpendicular to the magnetic field. This is crucial for identifying the processes responsible for the nonthermal broadening. In the case of MHD waves, for example, if a region where the magnetic field is perpendicular to the LOS shows larger nonthermal broadening, this can provide support for an interpretation in which transverse waves, such as Alfvénic waves, contribute significantly to the broadening. Conversely, a positive correlation of increased broadening for regions with magnetic fields parallel to the LOS direction can point to longitudinal waves, such as slow MHD waves, as a more likely process. Similarly, the superposition of unresolved field-aligned plasma motions (in a low-plasma-$\beta$ environment) also will be correlated with parallel LOS and magnetic fields.

For this reason, many previous studies have investigated the center-to-limb variation of the nonthermal broadening in solar transition-region lines emitted in the low corona. These studies generally assume that the magnetic field is typically oriented vertical to the solar surface, so the nonthermal broadening can be interpreted as motions parallel to the magnetic field when at the solar disk center and perpendicular to the field when at the limb. Most studies in the literature have reported that the broadening is not correlated with the distance from the disk center (R. Roussel-Dupre et al. 1979; J. T. Mariska 1992; K. Cho et al. 2023). This finding has been interpreted as being due to both transverse and longitudinal waves or field-aligned flows equally affecting the nonthermal broadening

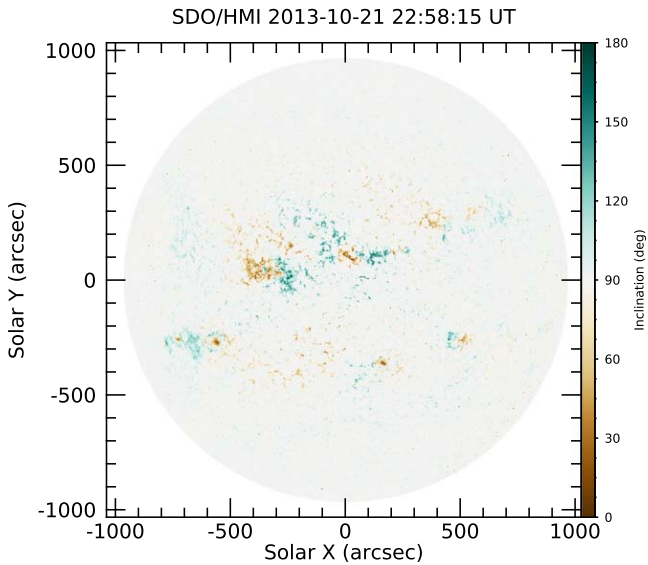

**Figure 1.** An example of a magnetic field inclination map from HMI observations. The white gray color (90°) indicates perpendicularity to the LOS direction.

(S. W. McIntosh et al. 2008; B. De Pontieu et al. 2015). Observations also show enhanced broadening near the solar limb (Y. K. Rao et al. 2022; M. Carlsson & B. De Pontieu 2023), above the solar limb (J. T. Mariska et al. 1979; S. W. McIntosh et al. 2008; M. Carlsson & B. De Pontieu 2023), and in the middle to upper corona (e.g., R. Esser et al. 1999; Y. Zhu et al. 2024). These results are used to support the claim that the transverse waves are a major source of the nonthermal broadening (M. Carlsson & B. De Pontieu 2023). At the same time, P. Testa et al. 2016 found that a comparison of disk and limb Interface Region Imaging Spectrograph (IRIS) observations of Fe XII in active region moss (i.e., in the transition region of hot loops) suggests the dominance of field-aligned motions, compatible with heating by magnetic reconnection. The literature suggests that a mix of both longitudinal and transverse effects appear likely to play a role. These effects may have a different relative importance for the observed broadening, depending on the type of solar region studied (quiet Sun or active region).

However, it is not clear that the magnetic field at transition-region heights is predominantly vertical. This has been studied for some numerical models of the quiet Sun (H. Peter et al. 2006), including the impact on the center-to-limb variation of line broadening, but for active regions, this has not been studied in detail. Yet, for such regions, it seems particularly likely that the magnetic field direction is not always vertical at transition-region heights, as it is dominated by the complex magnetic field topology caused by large-scale neighboring regions (plage and sunspots) of opposite polarity. The Helioseismic and Magnetic Imager (HMI; J. Schou et al. 2012) on board the Solar Dynamics Observatory (SDO; W. D. Pesnell et al. 2012) provides full-disk photospheric vector magnetograms at a 12 minutes cadence. The full-disk map of the inclination of the magnetic field with respect to the LOS (which we hereafter call the inclination) generated from HMI data does not show a clear center-to-limb variation (see Figure 1). This may be due to the photospheric magnetic field containing considerable horizontal field strength in the quiet Sun (D. Orozco Suárez 2012) or to the lack of sensitivity for measuring the transverse component in weak-field regions (M. G. Bobra et al. 2014). In addition, we need to consider the difference of the magnetic field inclination at photospheric heights and at transition-region heights, because the magnetic fields in the higher atmosphere should be determined by the configuration and proximity of neighboring magnetic polarities, even if most photospheric magnetic fields depart from the surface vertically. Therefore, it is worthwhile investigating the relation between nonthermal broadening and magnetic field inclination (with respect to the LOS) in strong enough magnetic field regions at the height of the transition-region emission, i.e., a few thousand kilometers above the surface.

In this study, we investigate the relation of the nonthermal broadening to the magnetic field inclination at transition-region heights in closed-field regions above plage regions. The plage regions are easily identified as bright regions in 1600 Å images observed with the Atmospheric Imaging Assembly (AIA; J. R. Lemen et al. 2012) on board SDO, which have a relatively strong magnetic field strength of greater than 50 G. We obtained Si IV 1403 Å spectra from IRIS (B. De Pontieu et al. 2014) full-disk mosaic data and performed Gaussian fitting to determine several spectral parameters. A nonlinear force-free field (NLFFF) extrapolation enables us to derive the magnetic field vector at transition-region heights. Through the comparison between the inclination angle of the magnetic field and LOS, and the spectral parameters in the transition region, we investigate the physical mechanisms driving nonthermal broadening in transition-region spectral lines.

## 2. Observation and Analysis

IRIS takes full-disk mosaic data at a monthly cadence.[6] Here, we use the data taken on 2013 October 21, when several well-developed active regions were distributed across the solar disk (see Figure 2), providing different viewing angles. The full-disk mosaic consists of 185 different pointings, and the field of view (FOV) of each pointing is about $128'' \times 180''$, with a spatial sampling of $2''$ and $0.66''$, respectively. The total number of pixels is $10{,}496 \times 548$ along the scanning and slit directions, and the total scanning time is about 14 hr. The IRIS mosaic data contain six spectral windows (Mg II h, Mg II k, Si IV 1394, 1403 Å C II 1334, and 1335 Å). Among them, we selected the Si IV 1403 Å line to investigate the nonthermal broadening in the transition region. The exposure time of the Si IV 1403 Å window was 2 s, with a spectral binning of 2.

We fitted the Si IV 1403 Å spectra with a single Gaussian function. To select spectra with sufficient statistics in the Si IV spectral emission line, to accurately derive its parameters, we only applied the single Gaussian fitting to the pixels where the peak of the Si IV was higher than $4 \, \mathrm{DN\, s^{-1}}$. The four parameters derived from the fitting are: Gaussian amplitude, Doppler shift, Gaussian width ($\sigma$), and background level (see Figure 3). Previous studies used a variety of definitions for spectral width, including Gaussian width, FWHM, and $1/e$ width; the relation among these parameters is as follows:

$$\mathrm{FWHM} = 2\sqrt{2\ln 2}\, \sigma, \quad w_{1/e} = \sqrt{2}\, \sigma. \qquad (1)$$

We chose the $1/e$ width for the nonthermal width definition, as this width (when expressed as a velocity) can be interpreted as the most probable velocity in a statistical sense. The nonthermal broadening is obtained by subtracting 1/e instrumental broadening ($w_{inst} = 0.023$ Å; B. De Pontieu et al. 2014) and 1/e thermal broadening, which corresponds to the Si IV peak ionization temperature of log $T$[K] ∼ 4.8 ($w_{th} = 0.011$ Å; B. De Pontieu et al. 2014), using the following relation:

$$w_{nth} = \sqrt{w_{1/e}^2 - w_{inst}^2 - w_{th}^2}. \qquad (2)$$

---

[6] https://iris.lmsal.com/mosaic_allin1.html

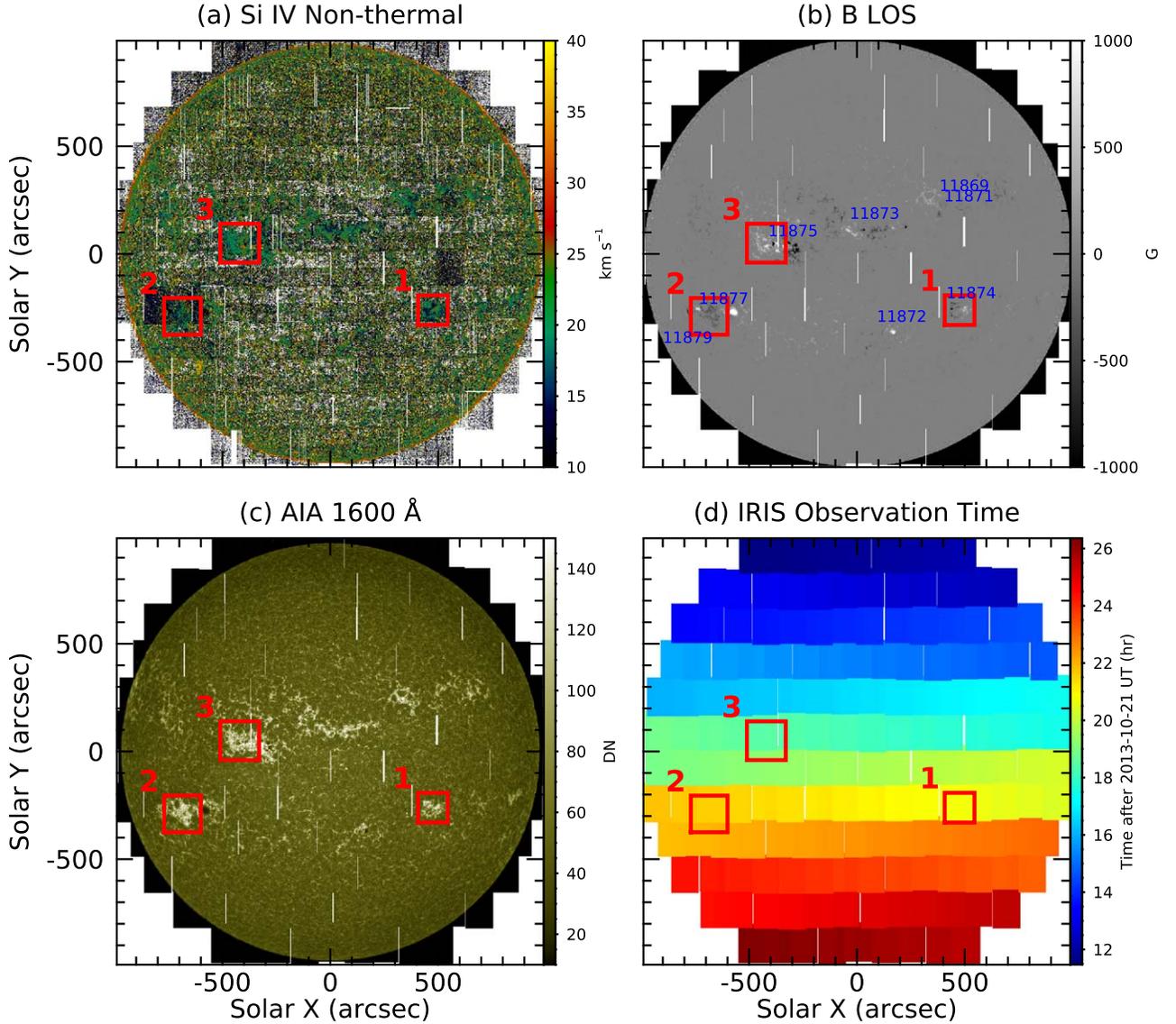

**Figure 2.** (a) Si IV nonthermal velocity map from IRIS mosaic observations. (b) LOS magnetic field mosaic map from HMI observations. The blue numbers indicate the locations of the numbered (NOAA) active regions. (c) AIA 1600 Å intensity mosaic map from AIA observations. (d) Observation time for the IRIS mosaic data. The red boxes indicate the selected rectangular areas or regions of interest. The white color indicates regions that lack data (e.g., because of data dropouts or because of gaps between different mosaic pointings). Note that the selected regions span across different pointings, and especially the time difference between two successive horizontal sequences is usually greater than 1 hr.

We assumed that the thermal broadening does not significantly vary with ionization temperature, because the response function of an Si IV ion has only a single narrow peak in the ionization equilibrium (see Figure 9 in H. Peter et al. 2022). So we fixed the thermal broadening regardless of the temperature variation.

The nonthermal broadening and the Doppler shift are converted to velocity units (Doppler velocity and nonthermal velocity). We only focus on the pixels that have a nonthermal velocity less than 40 km s$^{-1}$ and a Doppler velocity between ±30 km s$^{-1}$, to exclude poor fits for extremely noisy cases. We also generated the total intensity map of Si IV 1403 Å through integration of the spectra over a wavelength range of 1402.77 Å ± 1 Å.

To investigate the environments of our region of interest (ROI), we need to make mosaic maps from different observational data, analogous to the IRIS mosaic map. We generated AIA 1600 Å and HMI LOS magnetogram mosaic maps. Because of the long scan time for IRIS mosaic maps, the spatial (heliographic $x$ and $y$) position in these maps corresponds to different observing times. We selected the closest (in time) observational data of AIA and HMI for every IRIS slit position and filled their respective mosaic maps using 2D interpolation to allow pixel-by-pixel comparison.

We selected three rectangular regions of interest for this study. The red boxes in Figure 2 represent the selected areas. They are located at different positions on the solar disk and are parts of NOAA active regions 11874, 11877, and 11875, respectively (see Figure 2(b)). We considered each active region as a simple magnetic bipole and centered the rectangular area on the farther polarity region from the solar disk center between the two polarities. This choice is a way of maximizing the magnetic field inclination effect (Figure 4). If two polarities that are offset from the disk center are connected with potential-like magnetic fields, the inner edge (i.e., closer to the solar disk center) of the farther polarity region has parallel magnetic fields to the LOS direction at transition-region heights. On the contrary, the magnetic fields at the outer edge (i.e., closer to the solar limb) of the farther polarity region are perpendicular to the LOS direction. This effect is similar to the asymmetry of a sunspot's penumbra near the limb. The penumbra toward the center of the solar disk is narrower than that toward the solar limb. This is generally a geometric projection effect caused by the combination of the Wilson depression and the magnetic field inclination (A. Wilson & N. Maskelyne 1774). Meanwhile, the inclination effect illustrated in Figure 4 is not clear in the disk center region, so we did not include AR 11873 in our analysis.

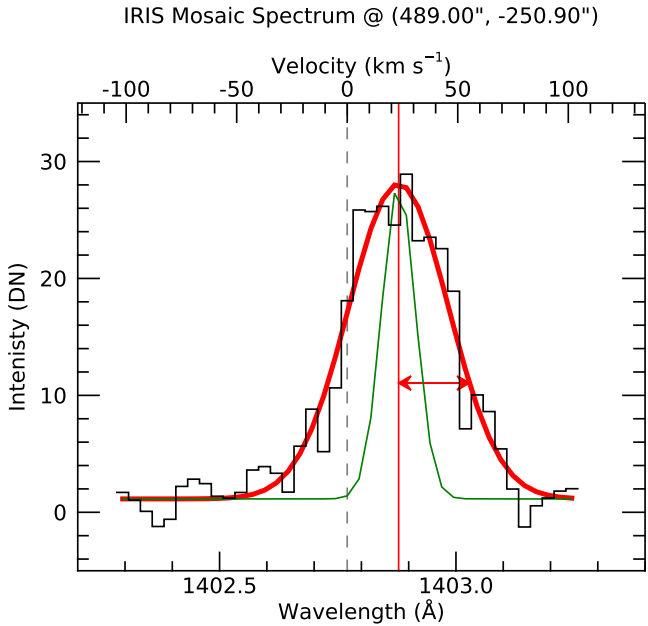

**Figure 3.** An example of an observed spectrum and a single Gaussian fit result. The location of the sampled spectrum is shown as the circles in Figure 6. The black histogram and gray dashed line indicate the observed spectrum and the rest wavelength of Si IV (1402.77 Å). The thick red solid line shows the result of a single Gaussian fitting. The red vertical line and red horizontal arrow show the Doppler shift (23.3 km s$^{-1}$) and $1/e$ width (31.1 km s$^{-1}$). The green line shows a spectrum with a thermal width corresponding to log $T$[K] = 4.8, the IRIS instrumental width, and a zero nonthermal width for comparison.

This regional dependence of the magnetic field inclination in our simple cartoon geometry can be validated using an NLFFF magnetic field extrapolation. We use the NLFFF package[7] in IDL SolarSoft for the NLFFF magnetic field extrapolation (M. S. Wheatland et al. 2000). For the extrapolation, the photospheric vector magnetic field at a specific moment is required as a boundary condition. We selected three different observation times when the IRIS slit passes through the center of the three rectangular areas and took the temporally closest HMI vector magnetograms as the boundary conditions. Since the local magnetic field extrapolation generally uses the

Cartesian coordinates, the result may have errors if we directly use the simply cropped observational data, because: (1) the physical size of a pixel depends on the location, due to the projection effect; and (2) the coordinates of the observed vector magnetogram differ from those of the local frame on the solar surface, except for a viewing geometry at the disk center. To minimize the projection effect, we converted the heliocentric Cartesian coordinates to spherical coordinates with the solar center as the origin and the solar radius as the distance. Then we made an equidistant grid, with its center located at the center of the ROI and a grid size of about 0°.03. The latter corresponds to the HMI pixel size (0″.5 ≃ 360 km at disk center). We selected a grid area that is sufficiently large to cover the whole active region that contains the rectangular area (Figure 5(a)). It seems that polarities outside of our grid area may affect magnetic fields near the edge of the active region. Although we think that our main interest, the variation of the inclination near the inner edge region, is not significantly affected by the coverage of the grid. To calculate the NLFFF in a local frame, we converted the magnetic field components from the observer's frame to a local frame that has solar east, north, and the normal of the solar surface as coordinate axes, then created input data using 2D interpolation (the underlying image in Figure 5(b)).

We obtained the 3D vector magnetic field through NLFFF extrapolation (Figure 5(b)). The calculated vector magnetic fields were converted to the observer's coordinates again (Figure 5(c)). We took the fields at 2200 km (≃7 pixels) above the photosphere as the transition-region height and calculated the angle $i$ between the extrapolated magnetic field direction at the transition region and the LOS direction (Figure 5(d)). We used the cos $i$ value for the analysis. The value cos $i$ = 1 indicates the magnetic field is perfectly aligned with the LOS direction. Finally, we generated a mosaic map of the magnetic field inclination at the transition region to allow pixel-by-pixel comparison with other data. There are various different NLFFF magnetic field extrapolation methods (see M. L. De Rosa et al. 2009). We expect that the result will not be very sensitive to the selection of the method if we use the same photospheric vector magnetic field as the boundary condition, because our target height, the transition region, is close to the input boundary.

For more accurate comparisons, we used two masks to define the ROIs in the rectangular areas. First, we confined our ROIs to the strong-magnetic-field region. This mask is important for two reasons: (1) it ensures that only regions in which the measurement error in the observed HMI vector magnetogram is less significant are included; and (2) it is more advantageous to study the effect of the magnetic field on nonthermal broadening than using weaker-magnetic-field regions. We perform a Gaussian smoothing in spatial dimensions with a $\sigma$ of 5 pixels, then take regions where the smoothed photospheric LOS magnetic field strength is stronger than ±50 G, depending on the sign of the main polarity. Second, we adopted a temporal masking. Since the NLFFF was calculated for the three specific times for each rectangular area, the observational data should be compared with the regions that were observed nearest in time, in order to reduce errors caused by temporal evolution or solar rotation. We chose a region that was observed within ±720 s, which corresponds to the time cadence of the HMI vector magnetogram, from the observing time of the input data for the NLFFF calculations. Since the selected rectangular areas span across different horizontal sequences of the IRIS mosaic (see Figure 2(d)), it is inevitable that the temporal masking removes some of the lower parts of the rectangular areas. The final mask combines both the strong-field mask and the temporal mask.

---
[7] https://sprg.ssl.berkeley.edu/~jimm/fff/optimization_fff.html

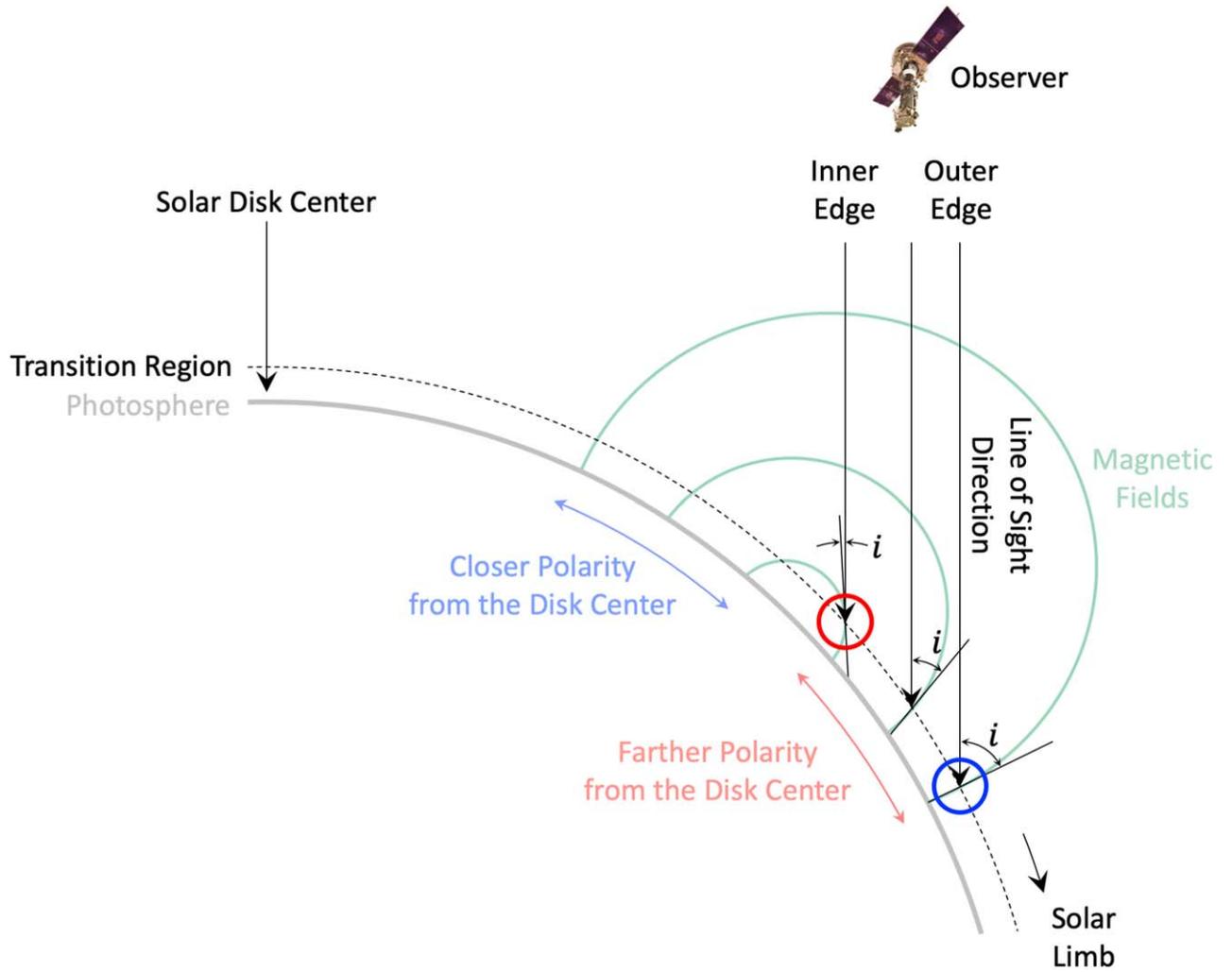

**Figure 4.** A schematic illustration of the magnetic field configuration and its relation to the LOS direction at transition-region heights. The definition of the inclination $i$ is described in several cases. If we assume that the active region located far from the disk center has a simple bipole configuration, the inclinations of the magnetic fields at the inner edge (i.e., toward disk center) of the farther polarity region are close to parallel to the LOS direction (the red circle). In contrast, the magnetic field inclination at the outer edge of the active region (i.e., toward the limb) is close to perpendicular to the LOS direction (the blue circle).

## 3. Results

Figure 6 shows the comprehensive information for region 1. It demonstrates that the region with a strong LOS field from the field strength masking (the blue contour in Figure 6(a)) matches the bright plage in the AIA 1600 Å intensity map (Figure 6(b)) and bright Si IV intensity region (Figure 6(c)). The AIA 1600 Å intensity in the ROI does not show a specific regional distribution. The magnetic field inclination, however, shows substantial variation. Figure 6(d) clearly shows that the magnetic field at transition-region heights in the northeastern part of the plage region is parallel to the LOS direction (red area), while the southwestern part is not (white or light purple area). Particularly, the northeastern direction is the direction to the solar disk center (the white arrow in Figure 6(d)). Considering that we chose the farther (from the disk center) polarity region as our ROI, the northeastern part becomes the inner edge of the farther polarity region. Thus, the magnetic fields are closer to parallel to the LOS in the inner edge of the farther polarity region, in agreement with the expectation from our cartoon (see Figure 4).

The most important finding is that the distribution of the Si IV nonthermal velocity also shows a similar pattern as the magnetic field inclination (Figure 6(e)). The averaged nonthermal velocity in the ROI is about $20\,\mathrm{km\,s^{-1}}$ (green color). One can find that the distribution of the nonthermal velocities is not uniform, but the pixels with higher nonthermal velocity are concentrated in the region toward the disk center. When comparing with the magnetic field inclination map, this region of enhanced broadening appears to be associated with the region where the magnetic field is aligned with the LOS direction. To further investigate and quantify this, we analyzed a scatter plot (pixel-by-pixel) between the two parameters. The 2D histogram in Figure 6(g) does indeed show an association between nonthermal velocity and the cosine of the magnetic field inclination. They have a positive linear correlation, with a Pearson correlation coefficient of 0.46. The linear fitting result demonstrates that the region with a magnetic field parallel to the LOS shows nonthermal velocities that are about $12.7\,\mathrm{km\,s^{-1}}$ higher than in the region with a magnetic field

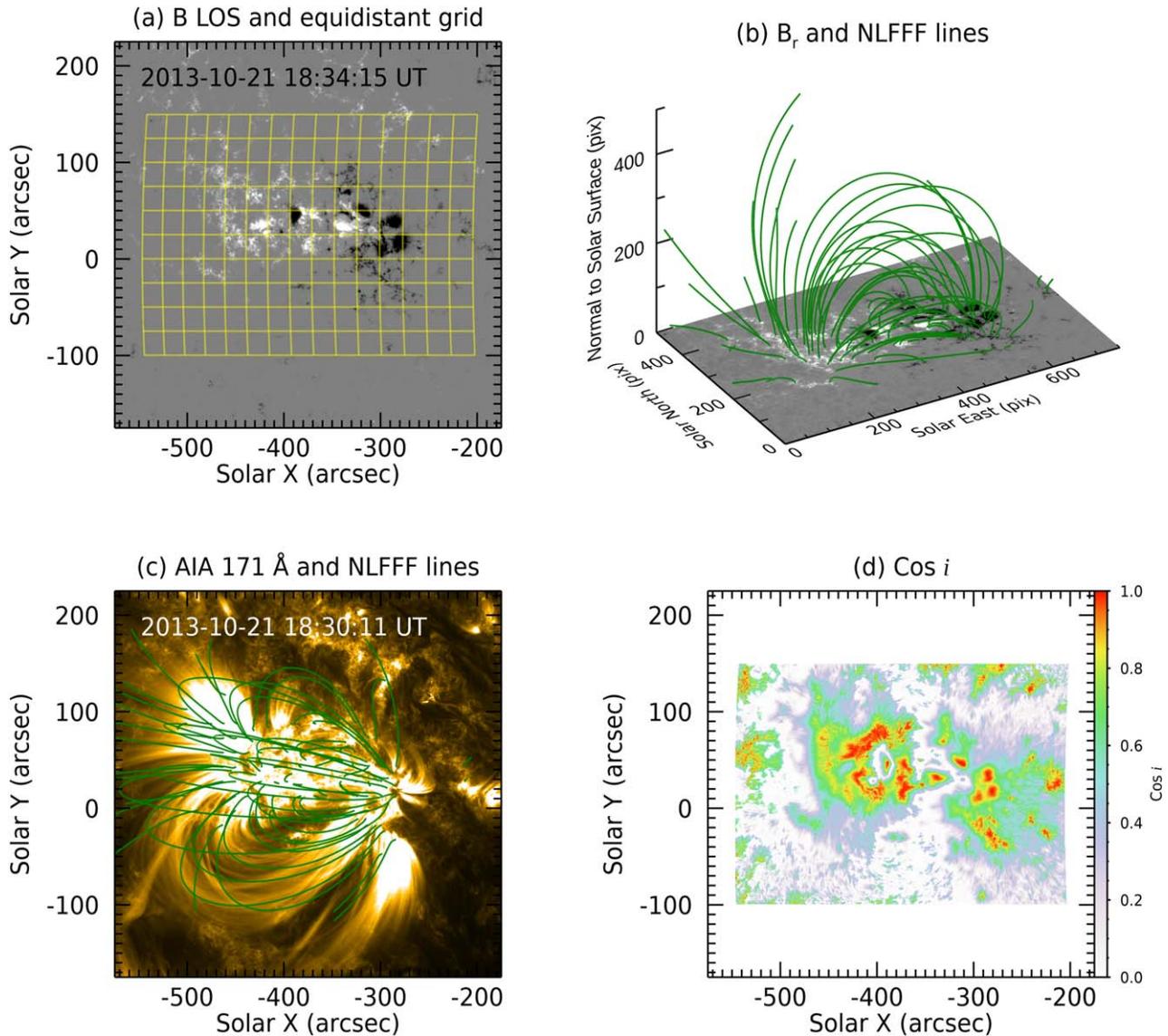

**Figure 5.** An example of how we obtain an inclination map for region 3 in Figure 2. (a) An equidistant grid is superimposed on the HMI LOS magnetic field map. The yellow solid lines indicate every 50 grid points. The observing time of the HMI data is shown at the top of the panel. (b) The NLFFF extrapolated magnetic fields (green lines) in the local frame. The underlying image shows the equidistant grid data of the radial magnetic field in the photosphere. (c) The NLFFF extrapolated magnetic fields (green lines) superimposed on the AIA 171 map. The observing time of the AIA data is shown at the top of the panel. Considering that the coronal loops in the 171 Å image outline the coronal magnetic field lines, one can consider the goodness of the magnetic field extrapolation. (d) The cosine of the magnetic field inclination at transition-region heights (about 2200 km above the photosphere). A value of 1 indicates magnetic field lines that are parallel with the LOS.

perpendicular to the LOS, i.e., the broadening is enhanced by a factor of 2.1.

The distribution of the Doppler velocity is also well associated with the cosine of the magnetic field inclination (Figure 6(f)). Relatively strong downflows ($\sim$20 km s$^{-1}$) were observed in the inner edge region. The 2D histogram between the cosine of the magnetic field inclination and the Doppler velocity similarly shows a positive correlation, with a Pearson correlation coefficient of 0.48 (Figure 6(h)). As both nonthermal and Doppler velocities show similar behavior with respect to the cosine of the magnetic field inclination, they also have a good correlation with one another. Figure 6(i) exhibits a positive correlation between them, with a Pearson correlation coefficient of 0.47. This latter result is consistent with a previous report (A. Ghosh et al. 2021). It is noteworthy that the Doppler velocities measured from the Si IV line generally show positive values, indicating that downflows are ubiquitous at transition-region heights in strong-field regions above plage.

The other ROIs also show similar results (Figures 7 and 8). Although their locations on the solar disk and viewing angles are different, they commonly exhibit, in the inner edge of the farther (from disk center) polarity, the following: a magnetic field parallel with the LOS, enhanced Si IV nonthermal velocity, and relatively strong downflow. The positive correlation between the parameters can also be found in the 2D histograms. The coefficients obtained from the 2D histograms are summarized in Table 1. The quantities show some regional variations. For example, in the case of the relation between the cosine of the magnetic field inclination and the nonthermal velocity, the Pearson correlation coefficients ($r$) and the gradients of the linear fitting ($c_1$) decrease in the order of regions 1, 2, and 3. This may be related to the complexity of the magnetic field configuration, i.e., deviation from the simple bipolar configuration assumed in our cartoon (Figure 4). Another notable difference is that the constant of the linear fitting between the magnetic field inclination and the Doppler velocity has a fairly large negative value in region 2 ($c_0 = -8.2$). We found that this was not real but was caused by an error in the wavelength calibration. Because the strong noise contaminates the reference line (O I 1355.5977 Å), the wavelength calibration has an error and the Si IV line looks blueshifted. It is responsible for the discontinuity at the boundary of the FOV in the velocity map in Figure 7(f) and the distinct coefficient $c_0$ in Figure 7(h).

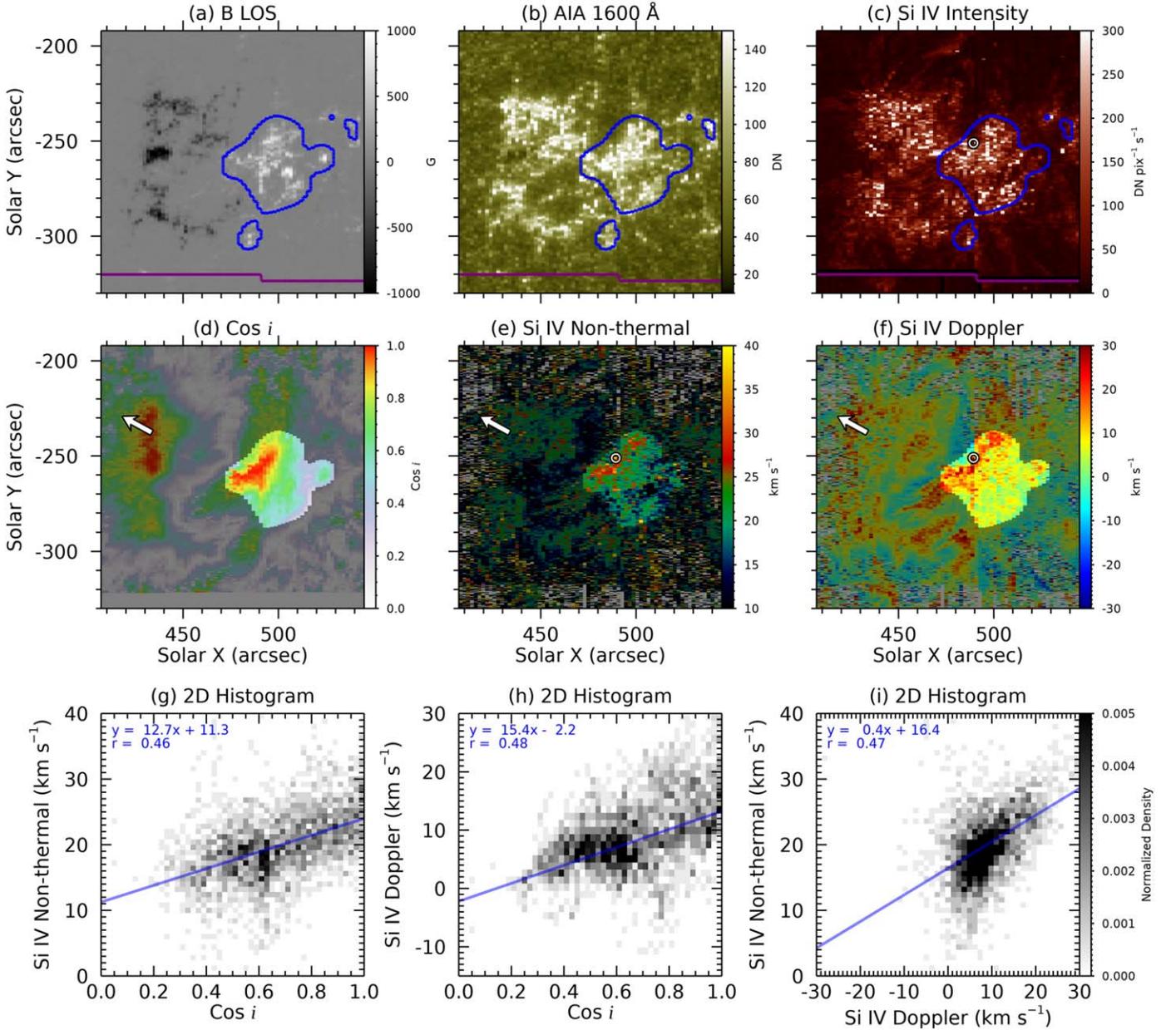

**Figure 6.** Observational data and 2D histogram for region 1: (a) LOS magnetic field strength map; (b) AIA 1600 Å intensity map; (c) Si IV intensity map; (d) map of the cosine of the inclination between the magnetic field at the transition region and the LOS, where the red regions, or $\cos i \simeq 1$, represent regions where the magnetic field is parallel to the LOS direction; (e) Si IV nonthermal velocity map; (f) Si IV Doppler velocity map; (g) 2D histogram between the cosine of the magnetic field inclination at the transition region and the Si IV nonthermal velocity; (h) 2D histogram between the cosine of the magnetic field inclination at the transition region and the Si IV Doppler velocity; and (i) 2D histogram between the Si IV Doppler velocity and Si IV nonthermal velocity. The blue contours and purple contours in (a)–(c) indicate the magnetic field strength masking and temporal masking, respectively. The unshaded areas in (d)–(f) indicate the major part of the intersection between the two maskings, the ROI. All analyses were conducted with the data within this unshaded area. The white arrows in (d)–(f) show the direction to the solar disk center. The linear fitting results are shown with the blue lines, and their coefficients and the Pearson correlation coefficients are exhibited in the upper left corner of every 2D histogram plot. The circles in (c), (e), and (f) indicate the positions of the sampled spectra in Figure 3.

### 4. Conclusion and Discussion

We have investigated the relation between the cosine of the magnetic field inclination, with respect to the LOS, and the observed Si IV nonthermal velocity. Our investigation is motivated by the fact that the inclination of the magnetic field with respect to the LOS is expected to have a different effect on the observed nonthermal velocity, depending on the main physical mechanism causing the nonthermal velocity. To find clues to the origins of transition-region nonthermal broadening, we compared the Si IV nonthermal velocity obtained from IRIS observations with the magnetic field inclination at transition-region heights from NLFFF extrapolations based on photospheric vector field measurements with HMI. We discovered that higher values of the Si IV nonthermal velocity are preferentially observed in regions where the magnetic field appears to be well aligned with the LOS direction. Additionally, relatively strong downflows also have a similar tendency as the nonthermal velocity. Our findings do not appear to depend on the relative location on the solar disk or different active regions. Our results hold true for region 2, where the data are significantly affected by noise introduced by cosmic rays.

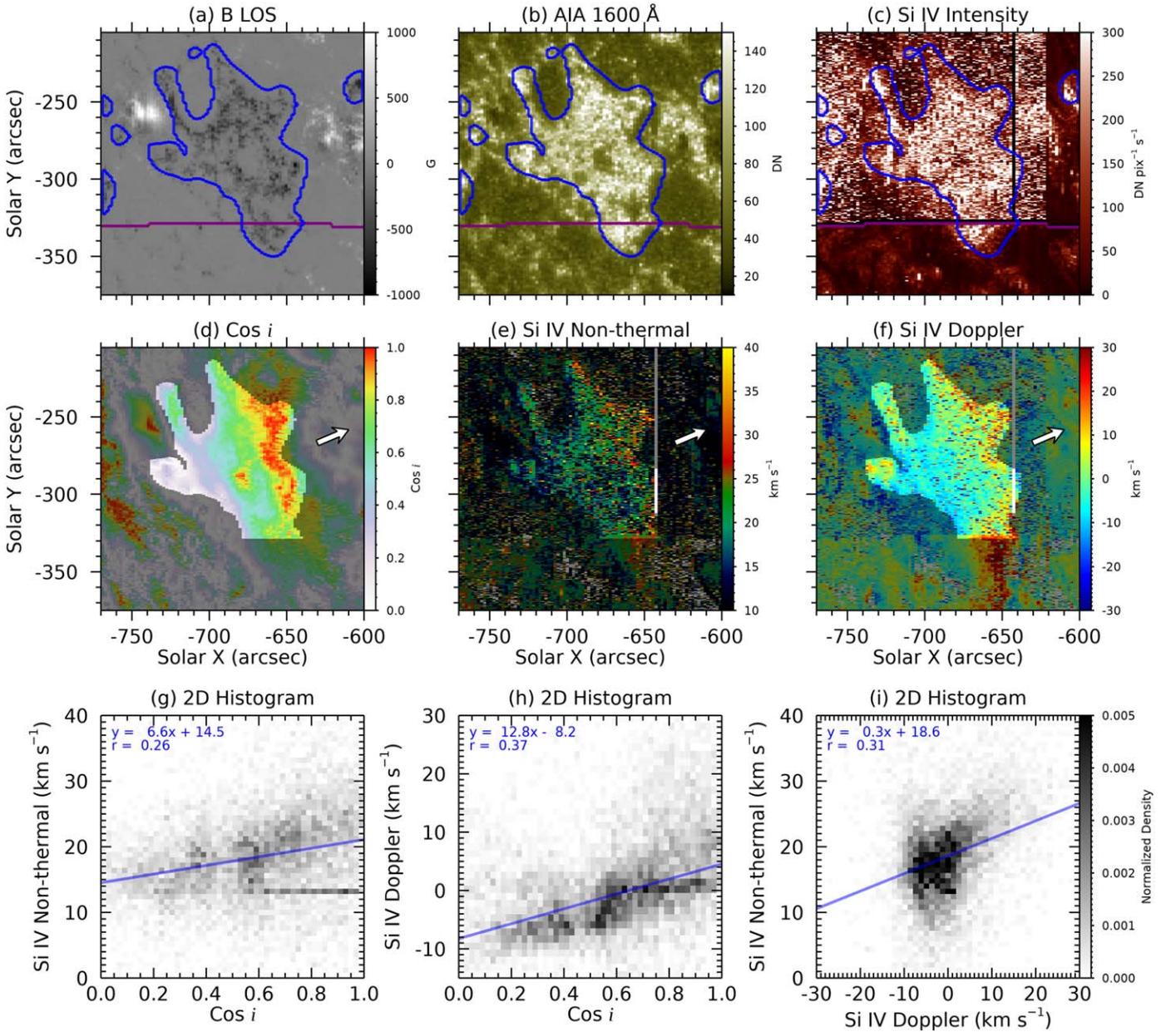

**Figure 7.** The same as Figure 6, but for region 2. The noisy patterns in the IRIS data were caused by cosmic-ray impacts on the detector, when IRIS passed through the South Atlantic Anomaly.

There is a possibility that our assumption of a fixed height for the Si IV line formation (when estimating the magnetic field inclination at transition-region height) introduces some uncertainty. However, we find similar correlations when we assume a fixed transition-region height of 4 Mm instead of 2 Mm, which indicates that our assumption of a fixed height does not significantly impact our conclusions.

Our findings imply that, in active regions, plasma motions along the magnetic field are a key contributing ingredient for causing enhanced nonthermal broadening in the transition region. These findings are in agreement with the analysis of the hotter transition-region plasma observed with IRIS in Fe XII emission (P. Testa et al. 2016), which also hinted at a dominance of field-aligned flows (although they did not estimate a pixel-by-pixel inclination of the magnetic field with

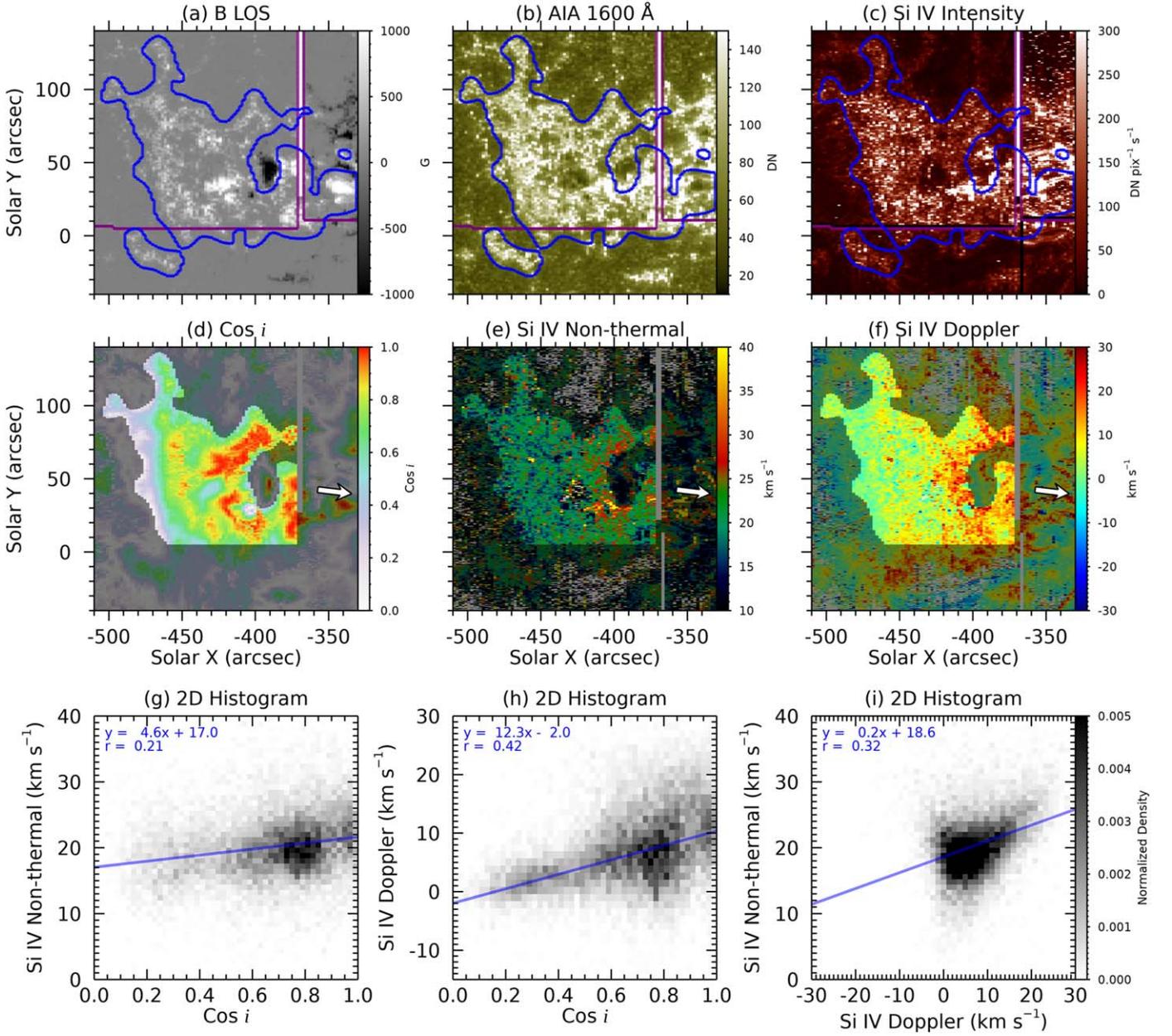

Figure 8. The same as Figure 6, but for region 3.

Table 1
Coefficients[a] from the Relation between Parameters and Information about the Active Regions

| Target | $B_{inc}$ versus $v_{nth}$[b] | | | $B_{inc}$ versus $v_{Dop}$[c] | | | $v_{Dop}$ versus $v_{nth}$[d] | | | Magnetic Classification[e] | $\mu = \cos\theta$[f] |
|---|---|---|---|---|---|---|---|---|---|---|---|
| | $c_1$ | $c_0$ | $r$ | $c_1$ | $c_0$ | $r$ | $c_1$ | $c_0$ | $r$ | | |
| Region 1 | 12.7 | 11.3 | 0.46 | 15.4 | −2.2 | 0.48 | 0.4 | 16.4 | 0.47 | $\beta$ | 0.57 |
| Region 2 | 6.6 | 14.5 | 0.26 | 12.8 | −8.2 | 0.37 | 0.3 | 18.6 | 0.31 | $\beta/\beta\gamma$ | 0.78 |
| Region 3 | 4.6 | 17.0 | 0.21 | 12.3 | −2.0 | 0.42 | 0.2 | 18.6 | 0.32 | $\beta\gamma/\beta\gamma\delta$ | 0.44 |

**Notes.**
[a] The linear fit coefficients in a form of $y = c_1 x + c_0$. The Pearson correlation coefficient is shown as $r$.
[b] Panel (g) in Figures 6, 7, and 8.
[c] Panel (h) in Figures 6, 7, and 8.
[d] Panel (i) in Figures 6, 7, and 8.
[e] http://helio.mssl.ucl.ac.uk/helio-vo/solar_activity/arstats-archive.
[f] Based on the center of the rectangular region.

respect to the line of sight, but instead assumed a dominance of the magnetic field orientation perpendicular to the surface, as in previous works). We find that regions where the magnetic field is parallel to the LOS have a nonthermal broadening of maximum 2.1 times higher than regions where the magnetic field is perpendicular. In the latter regions, however, the broadening is not zero. It thus appears that there are (at least) two different mechanisms at play. The mechanism we focus on here is the newly discovered field-aligned mechanism.

Which kind of physical mechanism can explain the observed relationship for these regions where the field is parallel to the LOS? Previous studies have traditionally suggested that MHD waves (e.g., Alfvénic waves or slow MHD waves) may be the dominant mechanism behind enhanced nonthermal velocities, as mentioned in Section 1. It seems clear that Alfvénic waves are not the major source responsible for the inclination dependence, at least in our cases. If the Alfvénic waves were the major contributor to the nonthermal velocity in the observed regions, we should detect an enhancement of nonthermal broadening in regions where the magnetic fields are perpendicular to the LOS direction, since Alfvénic waves are transverse waves. This is the opposite to what we see in our observations. Slow MHD waves could, in principle, be a candidate mechanism, since these would show a positive correlation between nonthermal velocity and the cosine of the inclination angle between the magnetic field and LOS. However, if slow-mode waves were the (only) dominant mechanism, one would expect alternating blue- and redshifts, so that an IRIS Doppler velocity map should be dominated, statistically, by a mix of blue- and redshifts. This is not the case in our observations, which are dominated by the well-known pervasive redshifts in the transition region (J. T. Mariska 1992; J. Chae et al. 1998; H. Peter & P. G. Judge 1999; V. H. Hansteen et al. 2010; H. Skogsrud et al. 2016). In other words, if slow-mode waves are to contribute to the nonthermal broadening, any explanation also needs to simultaneously explain the pervasive redshifts. It cannot be excluded that a mechanism causing pervasive redshifts and a separate mechanism (like slow-mode waves or shock waves; B. De Pontieu et al. 2015) are both acting together in combination at the same time.

An alternative explanation is that the nonthermal velocity is caused by a velocity gradient (within the height range over which the transition-region spectral lines form) associated with the downward field-aligned motions that we observe. It is clear that we are observing falling material, from the observed redshifts. Since we are observing regions with low plasma $\beta$, the plasma motions we observe are moving along the field. One potential scenario is that given the density stratification in the atmosphere, as plasma falls down into the lower atmosphere, it encounters increasing density, and eventually will slow down and stop falling at some height. During this process, a deceleration phase is inevitable, and this could lead to a change of velocity within the height range over which the spectral line is formed. This change in velocity could then lead to enhanced line broadening. This scenario is consistent with what we found in the observational data. There are several previous studies supporting our explanation. S. Patsourakos & J. A. Klimchuk (2006) simulated the responses of several spectra when nanoflares occur and reported that the spectra that formed just below 1 MK show weak redshift with a few tens of kilometers per second and broadening due to the mild draining motion. Even though these predictions do not extend to the low temperatures we observe here, and our result does not directly support the nanoflare occurrence, the characteristics of the observable parameters are roughly similar to the simulated results.

More generally, the well-known temperature dependence of the nonthermal velocity supports our notion. The nonthermal velocity has a peak around $3 \times 10^5$ K and decreases for increasing and decreasing temperatures (J. T. Mariska 1992; J. Chae et al. 1998). This signifies that, in closed-field regions on the solar disk, the nonthermal velocity is greatest at the transition-region temperature rather than at chromospheric or coronal temperatures (D. H. Brooks & H. P. Warren 2016). The density in the solar atmosphere is expected to change most dramatically in the transition region, so it is the region where steep density gradients, and thus strong velocity gradients, are most easily expected to occur. This also matches with our scenario. One caveat is that most previous studies of nonthermal line broadening did not distinguish between regions where the LOS is aligned or perpendicular with the magnetic field. Since it seems clear that there are at least two mechanisms at work, the selection of observations within the samples studied can have a significant effect on the interpretation of the results.

We note that many ions formed in the transition region and their associated spectral line emission are affected by nonequilibrium ionization effects. Such effects have been shown to significantly increase the temperature (and thus height) range over which the line is formed (K. Olluri et al. 2015). This could contribute to enhanced broadening as well.

In our observations, there are some differences between the observed regions. With respect to the Pearson correlation coefficients, region 1 shows a tighter linear correlation than region 3. We believe that the simplicity of the magnetic field configuration may play a role. For the NLFFF method, a simple configuration of magnetic poles as a boundary condition leads to a more realistic field extrapolation. We can see that region 1 is approximately close to a simple isolated bipole, so the NLFFF method can provide a more realistic estimate of the 3D field configuration. On the contrary, region 2 looks more complex and affected by another neighboring active region (AR 11879), and region 3 is even more complex, as evidenced by the occurrence of 10 C-class flares in this region on the observed date (see the magnetic classification in Table 1). Given the methodology used (NLFFF), it is thus perhaps not surprising that region 1 shows a higher correlation coefficient than the other regions. We have already mentioned that region 2 was observed by IRIS when the spacecraft passed through the South Atlantic Anomaly. This leads to more cosmic-ray hits from the impacts of energetic particles on the detector, causing noise in the obtained spectral parameters (See Figures 7(c), (e), and (f)). This can introduce noise in the observed correlations.

In summary, our results indicate that, in strong plage regions, there is a significant component to the nonthermal broadening that is field-aligned and associated with strong downflows. If we assume that the result from region 1 is an ideal case, there is still a considerable nonthermal velocity ($\sim$11.3 km s$^{-1}$) in the region where the LOS is perpendicular to the magnetic field (Figure 6(g)). This is not negligible, comparing with the field-aligned component ($\sim$12.7 km s$^{-1}$), and its origin is not well understood. We also see an enhancement of nonthermal velocity at the solar limb (see Figure 2(a)). As discussed, this may be a contribution from transverse waves, as pointed out by several previous studies (J. T. Mariska et al. 1978;

B. De Pontieu et al. 2015; Y. K. Rao et al. 2022; M. Carlsson & B. De Pontieu 2023). Our study has focused only on plage regions with strong magnetic fields. The nonthermal velocity in weak-field regions may be affected by different mechanisms, as examined by A. Ghosh et al. (2021). For example, it is not clear whether the magnetic field is strong enough to guide the plasma as well in weak-field regions. In addition, explosive events are (more) common in weak-field regions and likely are a significant contributor to the nonthermal velocity. We can even find signatures of explosive events in our data: a brightening is associated with high nonthermal velocity at $(-350'', 30'')$ in region 3 (Figures 8(b), (c), and (e)). Prevalent small events under the current detection limit may contribute to the nonthermal broadening in the same manner.

It thus seems likely that multiple processes cause nonthermal broadening of spectral lines. The relative importance of these processes may vary depending on the type of region. It is clear that revealing the origin of nonthermal broadening will give us valuable clues toward understanding the small-scale phenomena and energy transport in the solar atmosphere.

### Acknowledgments

We gratefully acknowledge support from the NASA contract NNG09FA40C (IRIS). P.T. was supported for this work by contract 8100002705 (IRIS) to the Smithsonian Astrophysical Observatory and by NASA grant 80NSSC20K1272. We gratefully acknowledge helpful discussions regarding the interpretation of the results with Viggo Hansteen and Juan Martinez Sykora. We greatly appreciate Alberto Sainz Dalda and Marc DeRosa, who helped with the use of the IRIS mosaic data and the NLFFF extrapolation, respectively. This research has made use of NASA's Astrophysics Data System and of the SolarSoft package for IDL. IRIS is a NASA small explorer mission developed and operated by LMSAL with mission operations executed at NASA Ames Research Center and major contributions to downlink communications funded by ESA and the Norwegian Space Agency (NOSA). Resources supporting this work were provided by the NASA High-End Computing (HEC) Program through the NASA Advanced Supercomputing (NAS) Division at Ames Research Center.

### ORCID iDs

Kyuhyoun Cho https://orcid.org/0000-0001-7460-725X
Bart De Pontieu https://orcid.org/0000-0002-8370-952X
Paola Testa https://orcid.org/0000-0002-0405-0668